\documentclass[preprint]{elsarticle}

\usepackage{hyperref}
\usepackage[nolist]{acronym}
\usepackage[onehalfspacing]{setspace}
\usepackage{booktabs}
\usepackage{float}
\usepackage{amsmath}
\usepackage{listings}

\journal{Nuclear Instruments and Methods A}

\begin{document}

\begin{frontmatter}

\title{Mirror Position Determination\\ for the Alignment of Cherenkov Telescopes}
\author[b]{J.~Adam}
\author[a]{M.~L.~Ahnen}
\author[b]{D.~Baack}
\author[c]{M.~Balbo}
\author[d]{M.~Bergmann}
\author[a]{A.~Biland}
\author[d]{M.~Blank}
\author[a,1]{T.~Bretz}
\author[b]{K.~A.~Bruegge}
\author[b]{J.~Buss}
\author[c]{A.~Dmytriiev}
\author[b]{M.~Domke}
\author[d,2]{D.~Dorner}
\author[b]{S.~Einecke}
\author[d]{C.~Hempfling}
\author[a]{D.~Hildebrand}
\author[a]{G.~Hughes}
\author[b]{L.~Linhoff}
\author[d]{K.~Mannheim}
\author[a]{S.~A.~Mueller\corref{mycorrespondingauthor}}
\author[a]{D.~Neise}
\author[c]{A.~Neronov}
\author[b]{M.~Noethe}
\author[d]{A.~Paravac}
\author[a]{F.~Pauss}
\author[b]{W.~Rhode}
\author[a]{A.~Shukla}
\author[b]{F.~Temme}
\author[b]{J.~Thaele}
\author[c]{R.~Walter} 
\cortext[mycorrespondingauthor]{Corresponding author: Sebastian Achim Mueller, sebmuell@phys.ethz.ch}
\address[a]{ETH Zurich, Institute for Particle Physics\\
Otto-Stern-Weg 5, 8093 Zurich, Switzerland}
\address[c]{University of Geneva,  ISDC Data Center for Astrophysics\\
 Chemin d'Ecogia 16,  1290 Versoix,  Switzerland}
\address[d]{Universit\"at W\"urzburg, Institute for Theoretical Physics and Astrophysics\\
Emil-Fischer-Str. 31, 97074 W\"urzburg,  Germany}
\address[b]{TU Dortmund, Experimental Physics 5\\
Otto-Hahn-Str. 4, 44221 Dortmund, Germany}
\address[1]{\scriptsize  also at RWTH Aachen}
\address[2]{\scriptsize  also at FAU Erlangen} 
%
\begin{abstract}
\acfp{iact} need imaging optics with large apertures to map the faint Cherenkov light emitted in extensive air showers onto their
image sensors. 
Segmented reflectors fulfill these needs using mass produced and light weight mirror facets. 
However, as the overall image is the sum of the individual mirror facet images, alignment is important. 
Here we present a method to determine the mirror facet positions on a segmented reflector in a very direct way.
Our method reconstructs the mirror facet positions from photographs and a laser distance meter measurement which goes from the center of the image sensor plane to the center of each mirror facet.
We use our method to both align the mirror facet positions and to feed the measured positions into our \ac{iact} simulation.
We demonstrate our implementation on the $4\,$m \acf{fact}.
\end{abstract}
\end{frontmatter}
\newcommand{\RadioCoeffCresponse}{r_\text{resp}}
\newcommand{\RadioCoeffTexp}{r_\text{expo}}
\newcommand{\RadioConstant}{r_\text{const}}
\newcommand{\NormalizedMirrorResponse}{R}
\newcommand{\StarIntensity}{s}
\newcommand{\MirrorReflectionIntensity}{m}
\newcommand{\RelativePointing}{\Theta}
\newcommand{\texp}{T_\text{expo}}
\newcommand{\iflux}{I_\text{pix}}
\newcommand{\cresponse}{C_\text{pix}}
\newcommand{\geom}{\alpha}
\newcommand{\FigCapLabSca}[4]{
    \begin{figure}[H]
        \begin{center}
            \includegraphics[width=#4\textwidth]{#1}
            \caption[]{#2}
            \label{#3}
        \end{center}
    \end{figure}
}
\newcommand{\FigCapLab}[3]{
    \FigCapLabSca{#1}{#2}{#3}{1.0}
}
\newcommand{\TwoFigsSideBySide}[2]{
    \begin{minipage}[t]{0.485\linewidth}
        \vspace{-0.5cm}
        \includegraphics[width=1\textwidth]{#1}
    \end{minipage}
    \hfill
    \begin{minipage}[t]{0.485\linewidth}
        \vspace{-0.5cm}
        \includegraphics[width=1\textwidth]{#2}
        \vspace{-1cm}
    \end{minipage}
}
\newcommand{\SideBySide}[2]{
    \newline
    \begin{minipage}[t]{0.485\linewidth}
        #1
        \end{minipage}
    \hfill
        \begin{minipage}[t]{0.485\linewidth}
        #2
    \end{minipage}\\
}
\section{Introduction}
\acfp{iact} have large effective areas to observe cosmic ray and gamma ray air showers and so have opened the very high energy gamma ray sky in the 1\,tera electron Volt regime to astronomy.
Almost \cite{CANGAROO1_optics} all former \cite{WHIPPLE_optics, CAT_Themis_optics}, current \cite{VERITAS_optics, HESS_I_optics, HESS_II_optics, MAGIC_optics, FACT_design}, and future \acsp{iact} \cite{CTA_Introduction, TAIGA_IACT_optics} make use of segmented imaging reflectors with apertures from about $10\,$m$^2$ up to several $100\,$m$^2$.
Segmented imaging reflectors with identical facets can be mass produced cost efficiently with an acceptable imaging quality which makes them a great choice for \acsp{iact}.
In \ac{iact} observations, characteristic spatial and temporal features, found in the recorded image sequences, separate the few gamma ray induced events from the much more numerous set of hadronic events.
However, manipulating the mirror facet orientations and positions to improve the image quality is a challenge.
Such alignment manipulations are not only important during the installation but also in case of repair and replacement of facets.
Here we present a method to measure and align the mirror facet positions with respect to the image sensor plane of the \ac{fact} \ac{iact} very directly and with only minimal modifications to the telescope.
Our mirror position determination tool can be an addition to an already existing mirror facet orientation alignment tool, to further reduce the Point Spread Function (PSF) area in the $5\%$ regime.
Beside its performance, our tool provides a well defined and simple procedure to maintain, refurbish, or characterize segmented imaging reflectors over their long life times.
%
\section{Motivation}
\label{secCurrentAlignmentMethods}
The first motivation is performance.
If one already has a mirror facet orientation alignment tool \cite{CAT_Themis_optics, FACT_Bokeh_alignment}, \cite{SCCAN_Arqueros_2005, VERITAS_SCCAN_alignment, FACT_NAMOD_alignment}, \cite{MAGIC_amc, HESS_II_optics}, and one wants to max out the performance of one's segmented imaging reflector, one might profit from a mirror facet position alignment system to gain $\approx 5\%$ in spatial and temporal resolution.
To quantify the performance impact, we simulate the PSF of \acs{fact} for first, a global offset of all mirror facets, and second a normally distributed spread of the individual mirror facet positions around their target positions.

The Figures \ref{FigGolbalZ}, \ref{FigStddevZ} and \ref{FigTimeVsZ} show the mean and the uncertainty of this mean for the two performance indicators PSF area and arrival time spread based on simulated distributions of 120 trails.
Here, only the mirror facet center positions along the optical axis $m_z$ are varied and the mirror facet orientations are always chosen accordingly to have the minimal PSF area which corresponds to an optimal orientation alignment.
The presented PSF area is not for on axis light, but for the more representative $1.5^\circ$ off axis region ($2/3$ of \acs{fact}'s field of view radius of $2.25^\circ$).
We choose the range for the offset and the spread to be up to $37\,$mm ($0.75\%$ of \acs{fact}'s focal length).
\FigCapLabSca{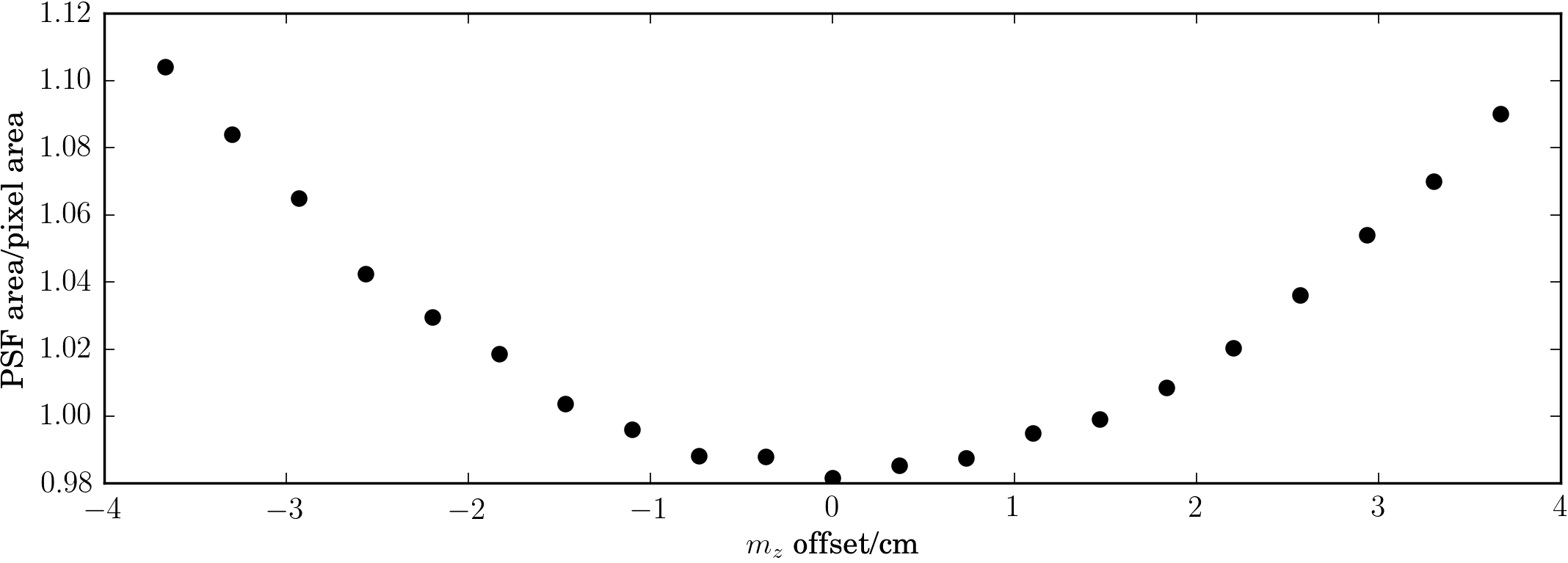}{
   Area of the \acs{fact} PSF against global offset of mirror facets along the optical axis. Note the supressed zero on the axis of the PSF area.
}{FigGolbalZ}{0.9}
\begin{minipage}[t]{0.46\linewidth}
    \vspace{-0.5cm}
    \begin{figure}[H]
        \begin{center}
            \includegraphics[width=1.0\textwidth]{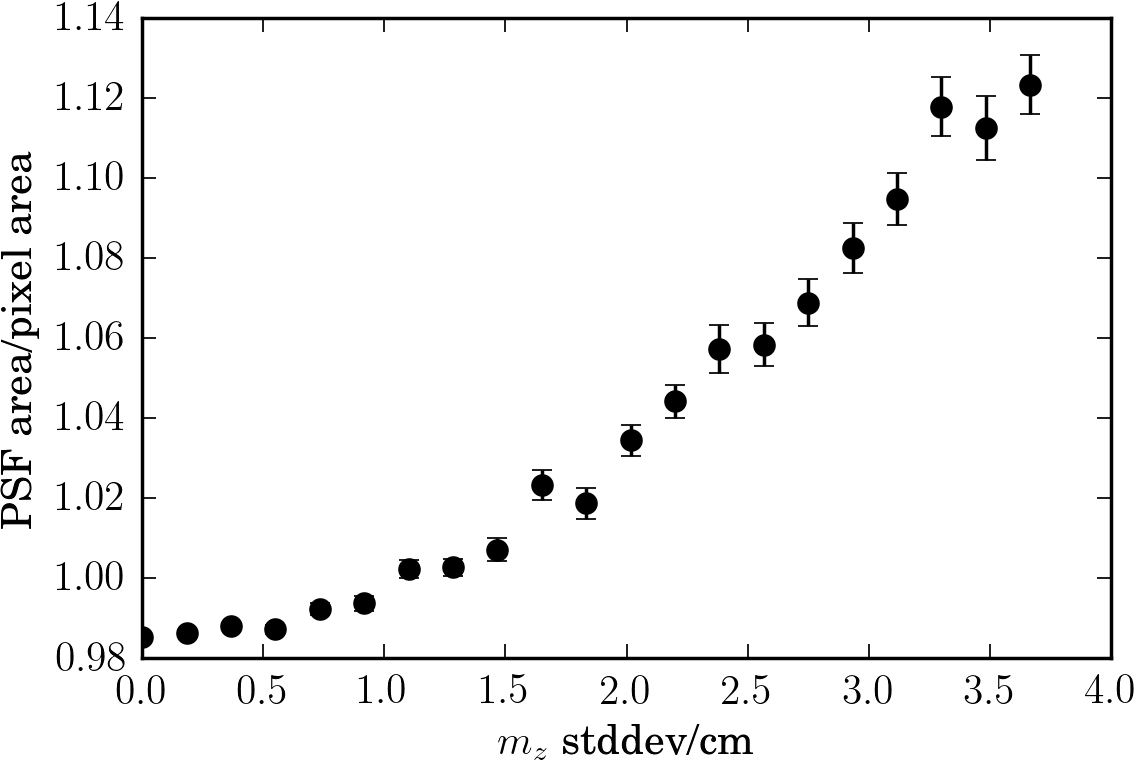}
            \caption[]{
                Area of the \acs{fact} PSF against the standard deviation of the mirror facet position spread along the optical axis.}
            \label{FigStddevZ}
        \end{center}
    \end{figure}
\end{minipage}
\hfill
\begin{minipage}[t]{0.46\linewidth}
    \vspace{-0.5cm}
    \begin{figure}[H]
        \begin{center}
            \includegraphics[width=1.0\textwidth]{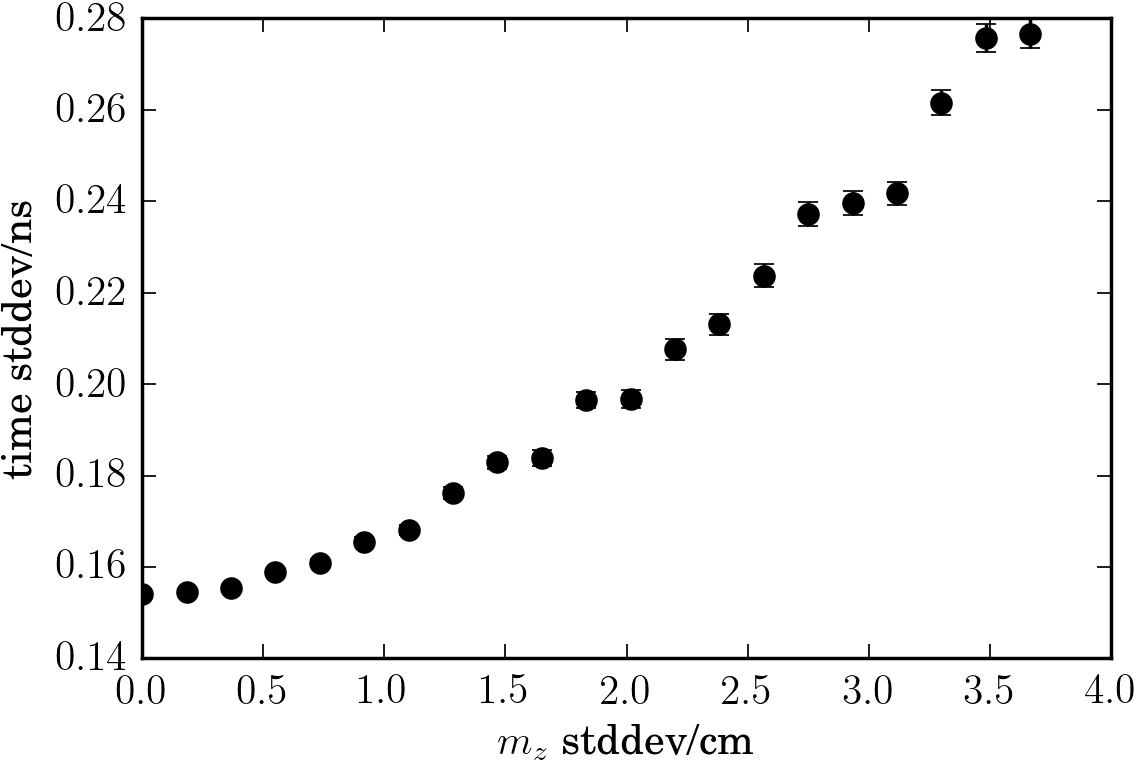}
            \caption[]{
                Standard deviation of the arrival time spread against the standard deviation of the mirror facet position spread along the optical axis.}
            \label{FigTimeVsZ}
        \end{center}
    \end{figure}
    \vspace{-0.5cm}
\end{minipage}
For \acs{fact}, we estimate from the simulations to have reduced the PSF area and the reflector time spread by $\approx 5\%$ due to our mirror facet fine positioning system.
The second motivation is to provide an easy and safe maintenance procedure.
For \ac{fact}, we want to make the most out of the vintage, but powerful components we inherited from the HEGRA \cite{HEGRA_status_and_results_1998} \ac{iact} array and so we designed the \acf{mipod} tool to fine align and measure the mirror facet positions in a reliable way while avoiding climbing on the reflector.
The task of mirror facet position alignment is usually neglected, because the optical support structure produces the mirror facet target positions well in the first place.
However, repair and upgrades will cause the mirror facet positions to deviate from their target positions over the life time of an \ac{iact}.
For example in the case of \acs{fact}, we modified the geometry of the segmented reflector in May 2014 from a pure Davies Cotton design \cite{Davies_Cotton_1957} towards a hybrid of Davies Cotton and parabola design to max out the timing resolution of \ac{fact}'s modern camera \cite{FACT_Muon_calibration_ICRC2015}.
In this process, all the mirror facet positions along the optical axis $m_z$ need adjustments, which vary from $+31\,$mm for some of the innermost mirror facets to $-21\,$mm for some of the outermost facets.
Our \acs{mipod} tool helped us to apply these unforeseen hardware changes to \acs{fact}'s refurbished HEGRA mount.\\
A third motivation for a mirror facet position determination is to measure the current state of a segmented reflector without modifying it to keep the \ac{iact} simulations up to date, and to investigate gravitational slump.
%
\section{Method and Implementation}
\label{SecApparatus}
In our \ac{mipod} implementation, we describe the positions of the mirror facets $\vec{m}$ with respect to the telescope's principal aperture plane.
First, we measure the mirror facet positions $m_x$ and $m_y$ perpendicular to the optical axis of the telescope using photographs of the segmented reflector's aperture.
Second, we calculate the mirror facet positions $m_z$ parallel to the optical axis by measuring the distances $w$ from the center of each mirror facet to the center of the image sensor plane with a remotely actuated \ac{ldm}.
%
\subsection{Mirror facet positions perpendicular to the optical axis, $m_x$ and $m_y$}
\label{SecPosXy}
We reconstruct $m_x$ and $m_y$ of a mirror facet in three steps using a photo camera to take pictures of the aperture of the telescope while the telescope and the photo camera are facing each other, see Figure \ref{FigFactAperture}.
First, we determine the scaling factor between distances in the picture with respect to distances on the segmented reflector by observing features in the picture which have well defined sizes, e.g. the mirror facet dimensions.
Second, we find the mirror facet positions in the picture by e.g. calculating the average of the mirror facet hexagonal corner positions. 
Third, we calculate the mirror facet positions $m_x$ and $m_y$ from the positions in the picture and the scaling factor.
\FigCapLabSca{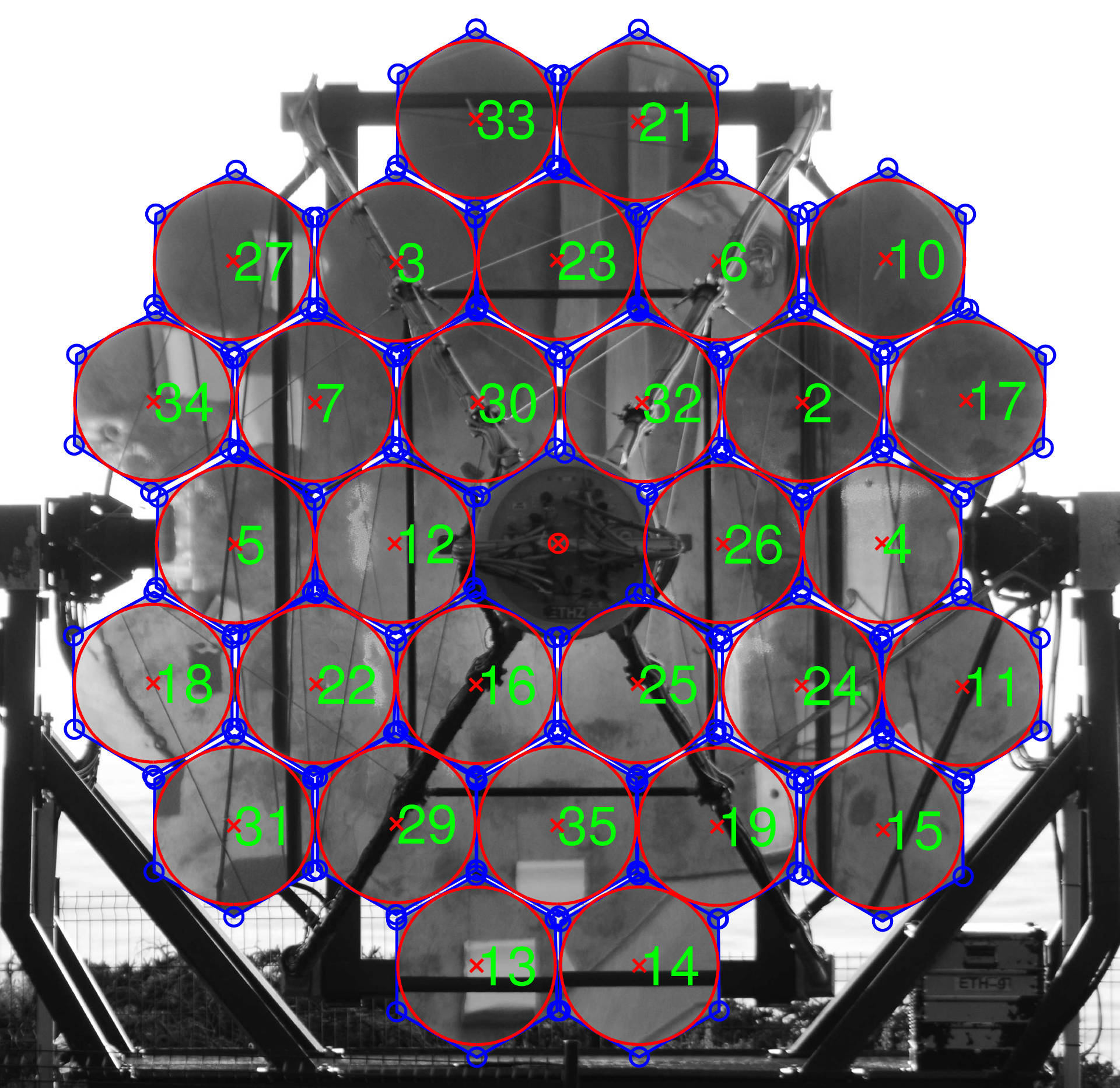}{
	Aperture picture to determine the mirror facet positions $m_x$ and $m_y$ perpendicular to the optical axis. 
	The blue circles and lines mark the edge of a mirror facet.
	The red crosses indicate the center of a mirror facet.
	In green, the IDs of \ac{fact}'s mirror facets are shown.
    This picture is also used for the Bokeh mirror orientation alignment \cite{FACT_Bokeh_alignment} for \ac{fact} and was taken using a digital single-lens reflex consumer camera.
}{FigFactAperture}{1.0}
The uncertainties $\Delta m_x$ and $\Delta m_y$ are estimated to be $\pm1\,$mm (0.03\% of the aperture diameter) using the spread across all the mirror facet edge lengths determined in the picture.
%
\subsection{Mirror facet positions parallel to the optical axis, $m_z$}
\label{SecPosZ}
We calculate the $i$-th mirror facet position ${m_i}_z$ parallel to the optical axis $\vec{z}$ by measuring the distance $w_i$ from the center of the $i$-th mirror facet to the center of the image sensor plane
\begin{eqnarray}
\label{EqMz}
{m_i}_z &=& d - \sqrt{w_i^2 - {m_i}_x^2 - {m_i}_y^2}.
\end{eqnarray}
Here, $d$ is the distance of the image sensor plane to the principal aperture plane, resulting from the reflector's focal length $f$ and the desired object distance to focus on.
\FigCapLabSca{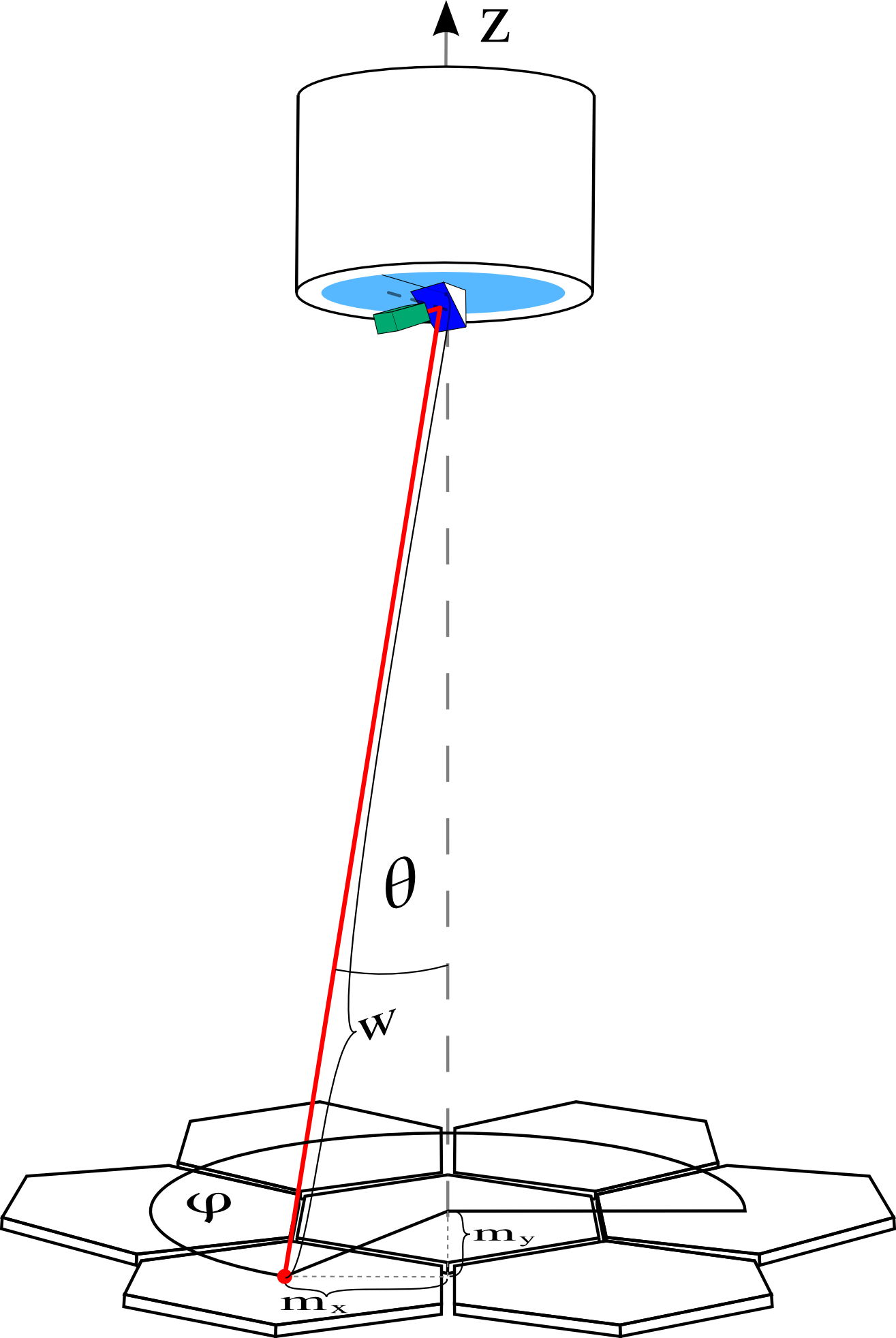}{
    The mirror facet positions $m_z$ are measured with respect to the telescope's image sensor plane using a remotely controlled \ac{ldm}, actuated in $\varphi$ and $\Theta$.
    The image sensor plane is shown in a light blue, the \ac{ldm} is shown in green and a $45^\circ$-mirror is shown in dark blue.
}{FigLdmOnFact}{0.685}
A \acf{ldm} measures the distances $w$ while being remotely guided from one mirror facet to the next one, see Figure \ref{FigLdmOnFact}.
To not touch the image sensor plane but to still measure $w$ as directly as possible, we were inspired by \cite{VERITAS_SCCAN_alignment} and place the \ac{ldm} on the virtual optical axis $\vec{z'}$ provided by a $45^\circ$-mirror on top of the image sensor plane of the telescope, see Figure \ref{FigPseudo}.
We say our \ac{mipod} measures the mirror facet positions rather directly because its mechanic defines the center of the image sensor plane as the main reference point.
This way, the number of measurements to determine a mirror facet position relative to the image sensor plane is lower than for methods which use arbitrary external reference points and therefore need to measure the position of the image sensor plane center additionally.
The direct measurement of $w$ further allows our \acs{mipod} to be practically used without a single line of software to interpret the intermediate measurements.
\FigCapLabSca{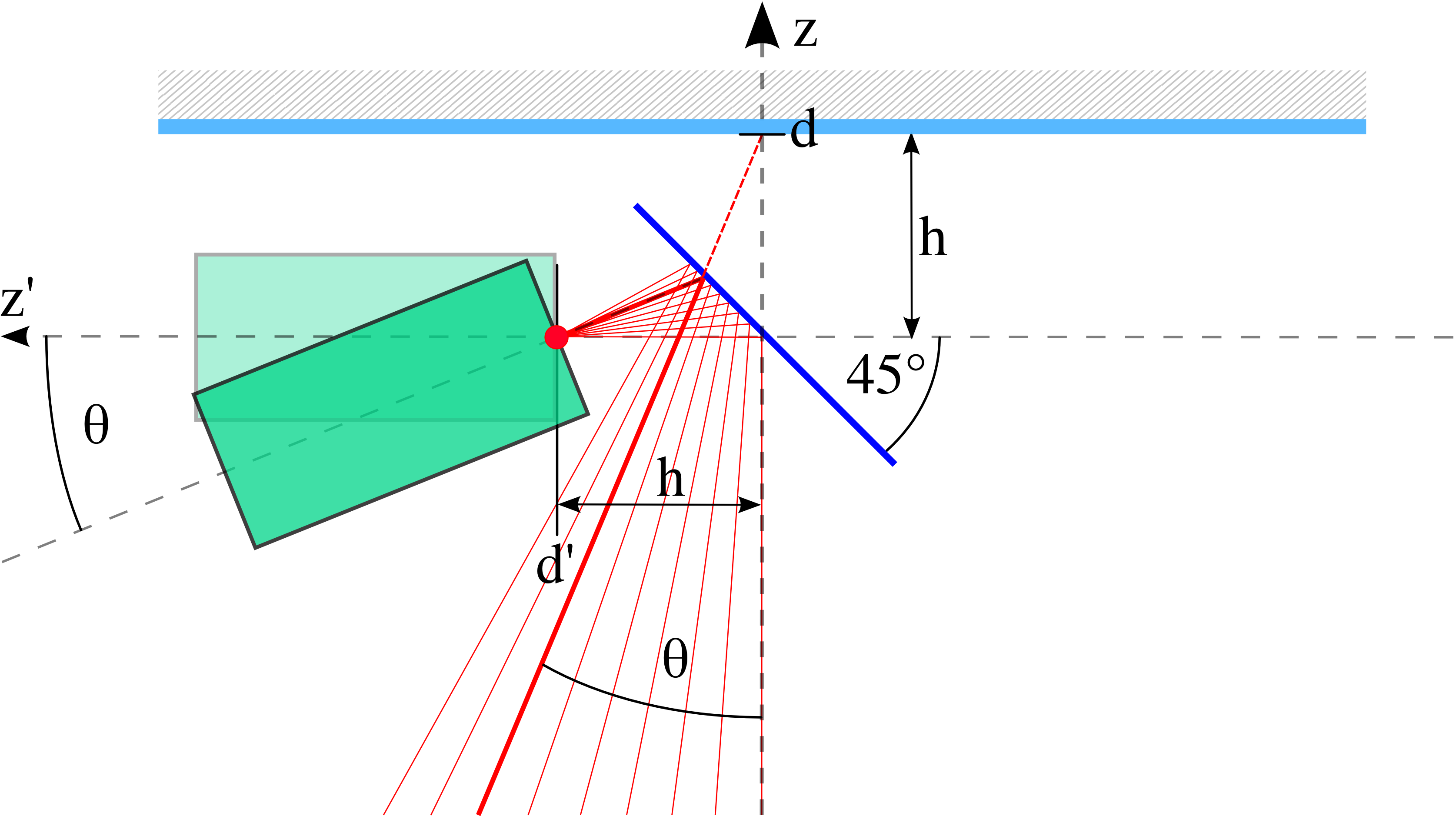}{
    Geometry of our \ac{ldm} mounting.
    The \ac{ldm} is shown in green, the image sensor of the telescope is shown in light blue and the $45^\circ$-mirror is colored in dark blue.
    The $45^\circ$-mirror provides the virtual optical axis $\vec{z'}$ in a distance $h$ above the telescope's image sensor plane.
}{FigPseudo}{1.0}
To point the laser beam of the \ac{ldm} to all the mirror facets, we implemented two consecutive actuated joints to rotate the \ac{ldm} in $\varphi$ and $\Theta$.
The rotation axes of the joints are chosen such that the virtual emission point of the \acs{ldm} is always in the center of the image sensor plane of the telescope, see Figure \ref{FigPseudo}.
The first actuated joint rotates the $45^\circ$-mirror together with the second actuator and the \ac{ldm} in $\varphi$ around the optical axis of the telescope.
The second actuator tilts the \ac{ldm} in $\theta$ around the virtual image sensor center $d'$ on $\vec{z'}$. 
The two axis joint can reliably move the beam of the \ac{ldm} on spots only $0.06\,^\circ$ in diameter ($5\,$mm respectively when mounted on \acs{fact}).
Based on cross check measurements in the lab, we estimate the uncertainty $\Delta w$ of the \acs{ldm} together with the actuated joints to be $\pm 1\,$mm for typical distances on \ac{fact} ($\approx 5\,$m).
%
\subsection{Slim optical table with $\varphi$ and  $\Theta$-joint}
\label{SecSlimGimbalJoint}
For a compact and lightweight mounting of the \ac{ldm}, and therefore a reliable measurement of $w$, the distance $h$ from the image sensor plane to the virtual optical axis should be small, see Figure \ref{FigPseudo}.
If $h$ is too large, the \ac{ldm} mounting position might be too distant from the optical axis $\vec{z}$, and might conflict with other components of the camera in order for the \ac{ldm} to freely rotate in $\varphi$ and $\Theta$.
In our implementation we approach this goal with two design features of the optical table which connects the remotely actuated \ac{ldm} to the telescope.
First, our optical table has the $\varphi$-joint built right into it.
Second, we use permanent magnets to mount components on the optical table instead of clamping plates with shape connectors, which reduces the thickness of the optical table down to only $6\,$mm.
Figure \ref{FigOpticalTable} shows the optical table used in our \ac{mipod} implementation and Figure \ref{FigOpticalTableScematics} shows a section view of it when mounted on a telescope.
\FigCapLabSca{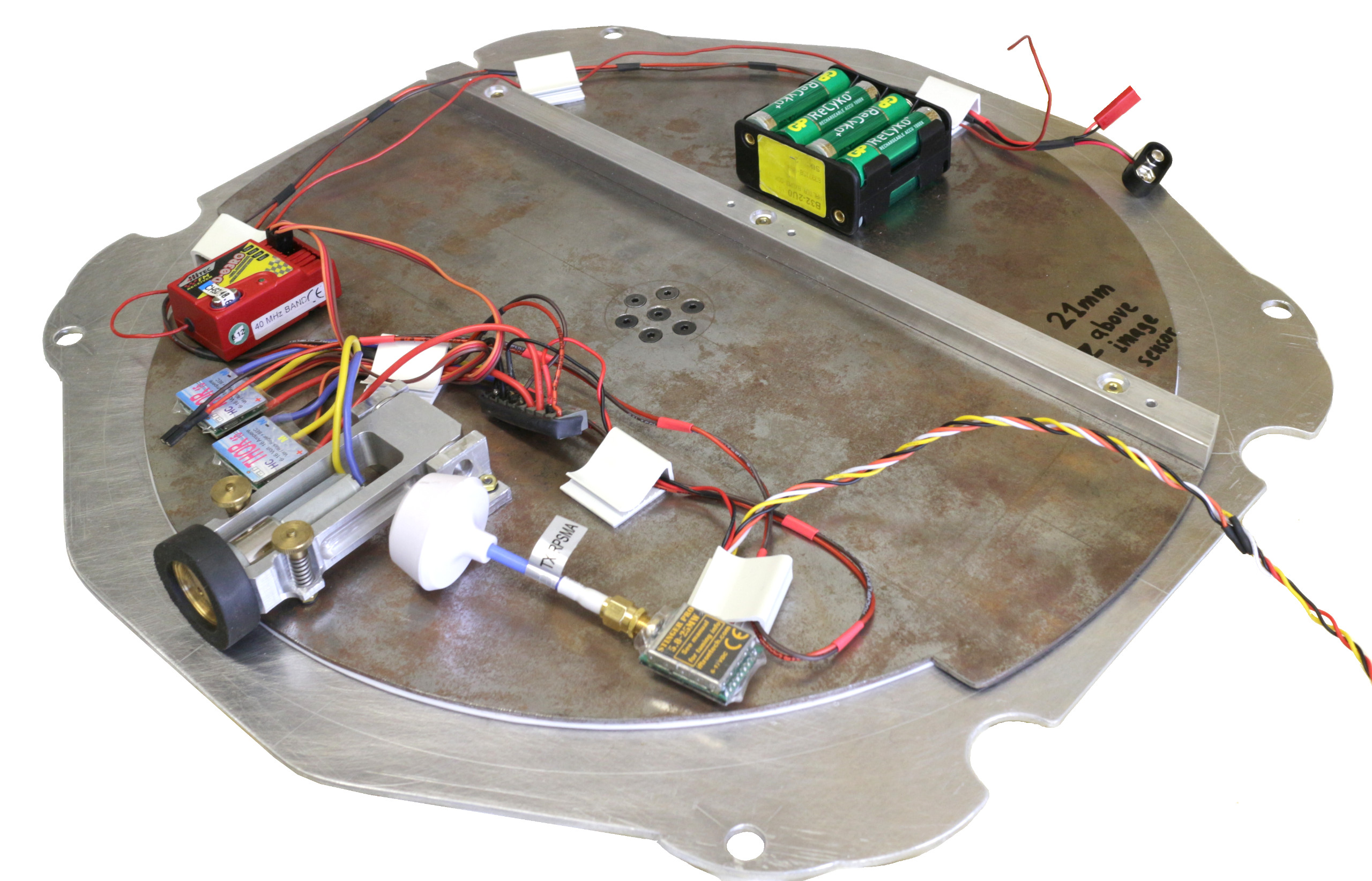}{%
    The optical table used to mount the remotely actuated \ac{ldm} on \acs{fact}.
    The rubber wheel actuator in the lower left part rotates the table.
    We used the same optical table to provide the virtual optical axis for our \acf{namod} for \acs{fact} \cite{FACT_NAMOD_alignment}. 
}{FigOpticalTable}{1.0}
\FigCapLabSca{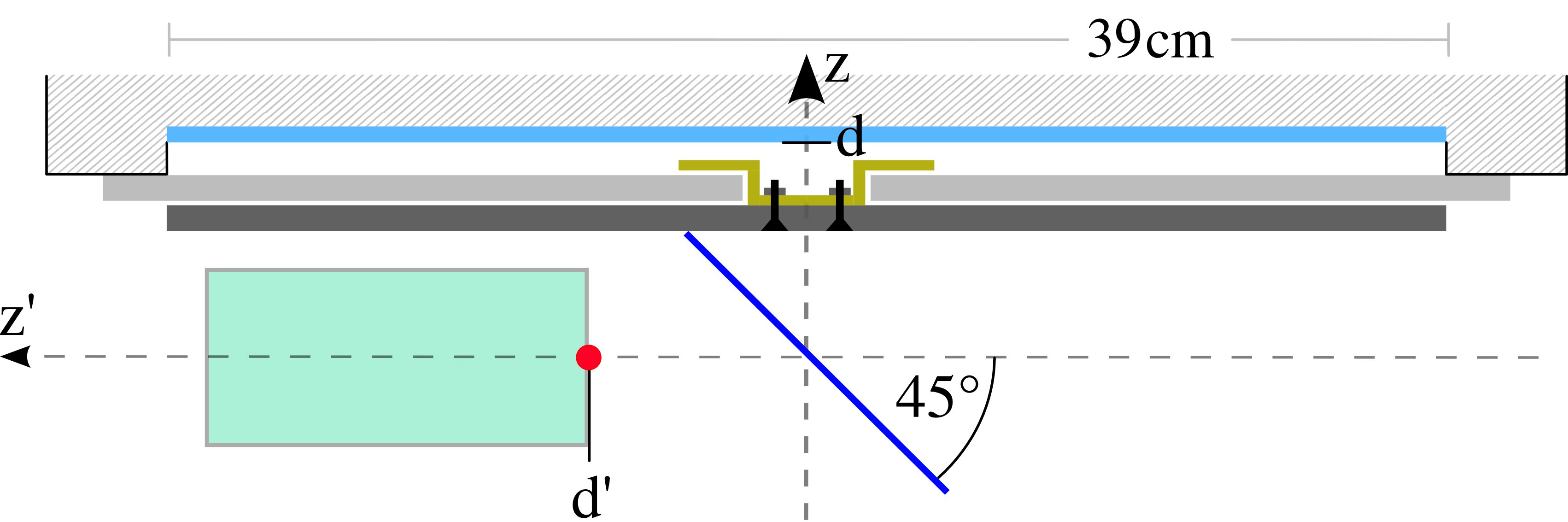}{%
    A section view of our slim optical table with its $\varphi$-joint along the optical axis $\vec{z}$ of the telescope.
    The image sensor of the telescope is shown in light blue and the image sensor housing volume is shown in hatched.
    Our optical table's alloy plate is shown in bright gray and the steel plate on top of it is shown in dark gray. 
    The brass friction joint is shown in yellow and mounted to the steel plate with bolts indicated in black.
    Here $d$ indicates the image sensor distance to the principal aperture plane along the optical axis $\vec{z}$ and $d'$ indicates the \ac{ldm} distance to the pseudo principal aperture plane along $\vec{z'}$.
}{FigOpticalTableScematics}{1.0}
The optical table is made out of a $3\,$mm alloy plate, which bolts on top of \acs{fact}'s image sensor, and a $2\,$mm steel disc on top of this.
The surface of the optical table is only $21\,$mm above the image sensor plane and has a profile rail to align its components.
The pseudo optical axis $\vec{z'}$ is $h=40\,$mm above the image sensor plane ($10.3\%$ of \ac{fact}'s image sensor diameter).
The $\varphi$-joint between the plates is actuated by a motorized rubber wheel which is mounted on the top steel plate while running on the bottom alloy plate, see Figure \ref{FigOpticalTable}.
\newline
The second actuated $\Theta$-joint of the \ac{ldm} is made out of a lightweight 3D printed plastic enclosure for the \ac{ldm} which includes a small actuator gear running along on a big 3D printed arc, see Figure \ref{FigThetaJoint}.
\FigCapLabSca{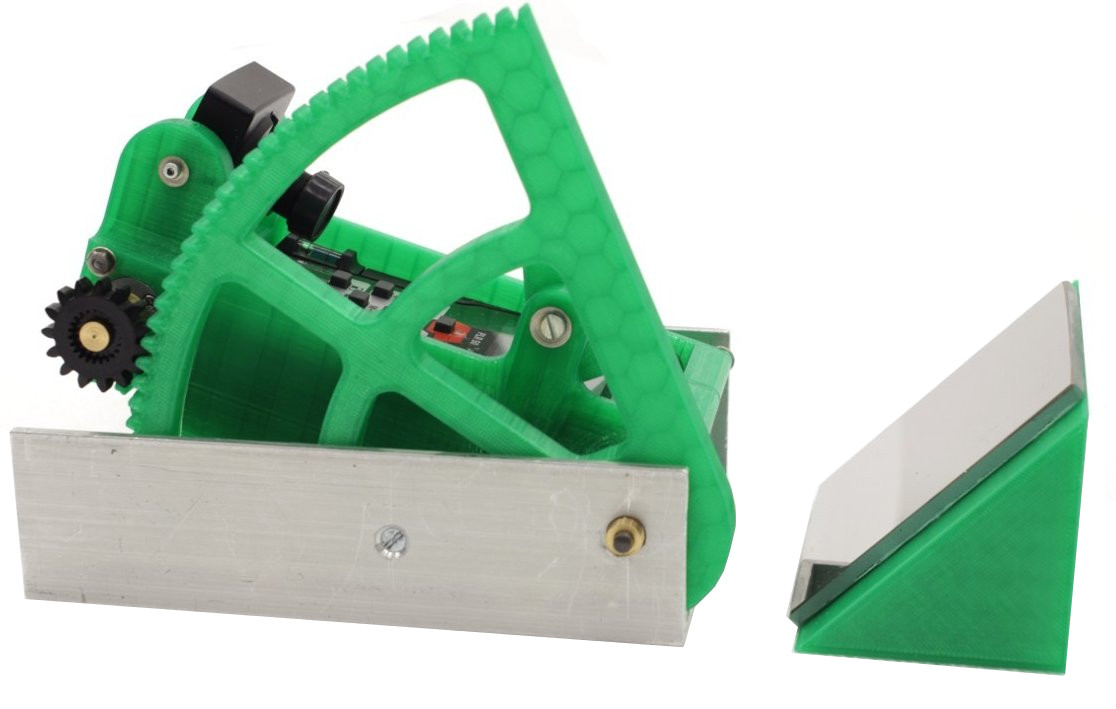}{%
    The 3D printed $\Theta$-joint of the \ac{ldm} and the front coated $45^\circ$-mirror.
    The two pieces are positioned on the optical table using permanent magnets.   
}{FigThetaJoint}{1.0}
%
\section{\ac{mipod} alignment for FACT}
\label{SecAlignmentForFact}
The \acf{fact} is located on Canary island La Palma, Spain. It inherited its mount and the mirror facets from HEGRA, see Figure \ref{FigLdmRedDot}.
While pioneering silicon photomultipliers for \acp{iact}, \acs{fact} is monitoring gamma ray bright blazars such as Mrk\,421 and Mrk\,501.
\ac{fact} has a focal length $f$ of $4.889\,$m and 30 identical, hexagonally shaped, and spherically curved mirror facets.
The maximum outer aperture diameter is $3.926\,$m.
Its image sensor spans $4.5^\circ$ field of view, or $39\,$cm in diameter respectively and has 1440 pixels which span $0.11^\circ$ or $9.5\,$mm each.
Typical air-showers records contain about $30\,$ photon equivalents spread over about $10\,$ shower pixels.
\FigCapLabSca{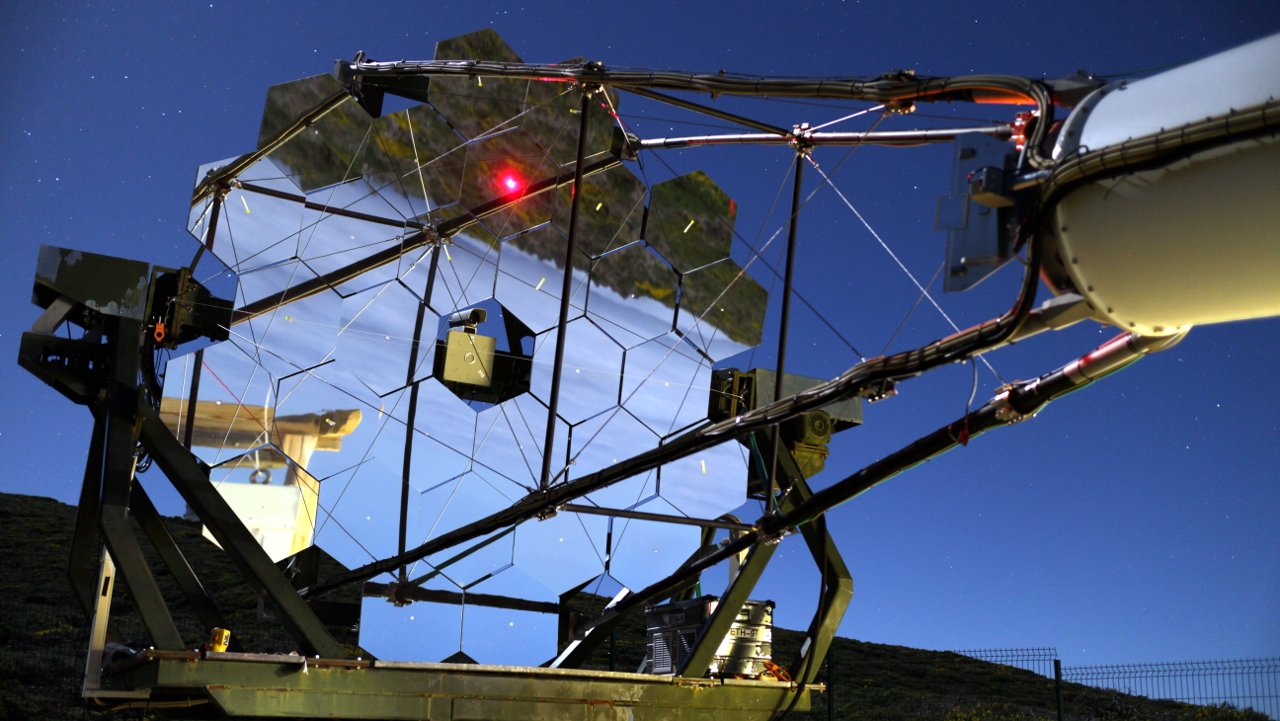}{
	The segmented reflector of \acs{fact} during our \ac{mipod} alignment. 
    The red laser beam of the \acs{ldm} is pointing to the central mirror facet in the second row from the top.
}{FigLdmRedDot}{1.0}
To redesign the \acs{fact} reflector with a hybrid Davies Cotton and parabola geometry, we use our \ac{mipod} alignment.
Further, we use \ac{mipod} to measure the positions of the mirror facets before and after the modification so that \acs{fact}'s \acs{iact} simulation is up to date.
%
\subsection{Installation and operation}
The optical table of our \ac{mipod} implementation is temporarily bolted on top of the image sensor of \acs{fact}, see Figure \ref{FigLdmInstalled}.
While observing the laser beam moving along the mirror facet surfaces, Figure \ref{FigLdmRedDot}, the \ac{mipod} operator controls the $\varphi$ and $\Theta$ actuators of the \acs{ldm} via an analog radio remote control.
The distance $w$ is measured continuously and displayed to the operator on a monitor.
Because of a former refurbishing process, most of the mirror facets on \ac{fact} have a dull spot of about $1\,$cm in diameter in the center of their surfaces which helped the \ac{ldm} to receive its scattered beam.
On the mirror facets without such dull spots, we attached dull stickers in the centers to run the \ac{ldm} reliably.
\FigCapLabSca{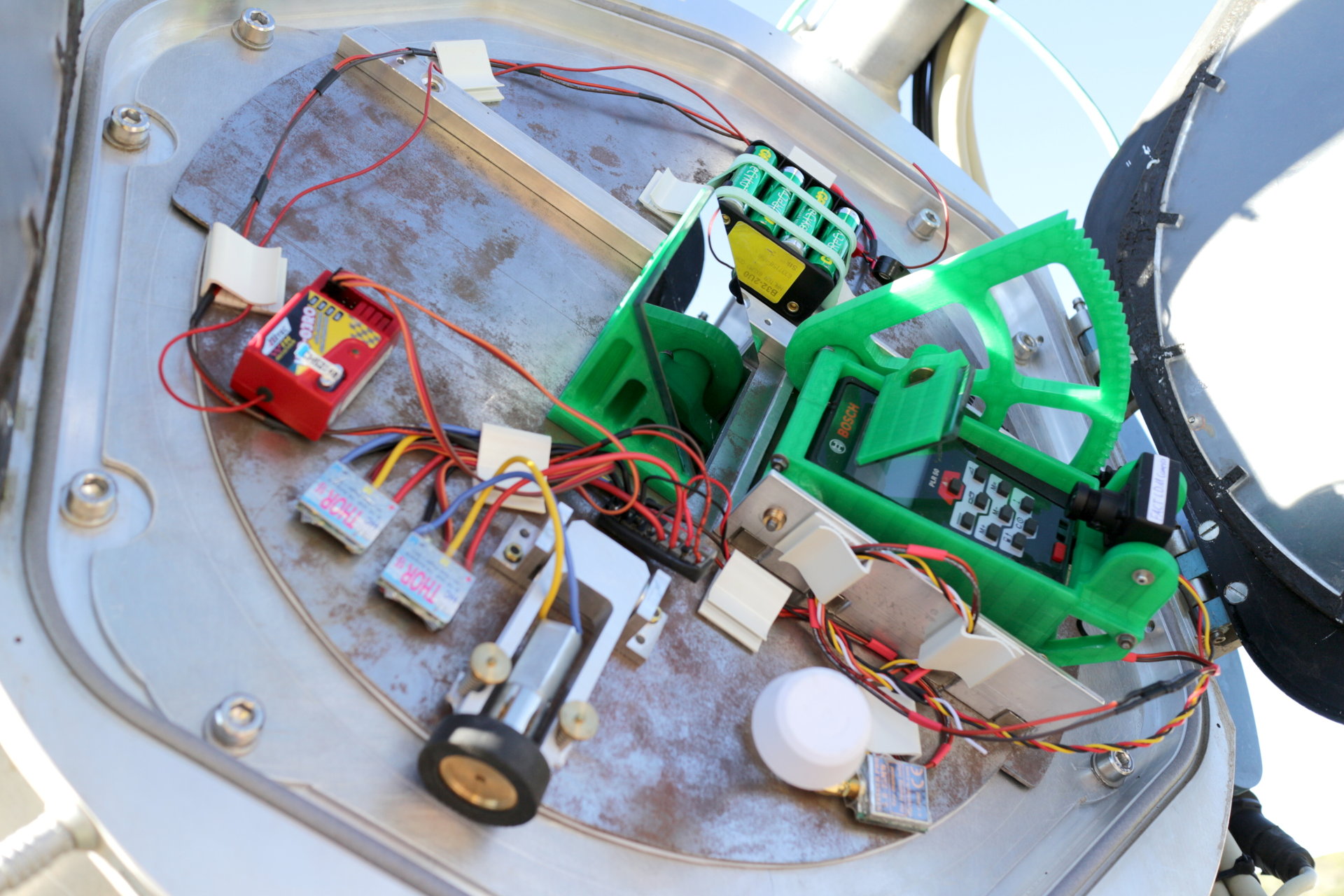}{
	Our full \ac{mipod} implementation mounted on top of the image sensor of \acs{fact}.
}{FigLdmInstalled}{1.0}
%
\section{Results}
\label{SecFactAlignmentResult}
First, our \ac{mipod} implementation measured that the mirror facets positions $m_z$ deviated by up to $-1.9\,$cm and $+2.6\,$cm (0.5\% of the  focal length $f$) from their target positions for the Davies Cotton geometry.
This deviation is now taken into account in our \acs{iact} simulation for this epoch.
Second, we aligned all the mirror facet positions in $m_z$ to their new target positions to form an equally mixed Davies Cotton and parabola hybrid geometry.
%
\section{Conclusion}
\label{SecConclusion}
Our \acf{mipod} implementation enabled us to reliably redesign the segmented imaging reflector of \acs{fact} by measuring the facet positions in a very direct way relative to the image sensor plane.
The optical table proved to be versatile as we used it also for our \acf{namod} \cite{FACT_NAMOD_alignment}.
Operating a vintage telescope mount and refurbished mirror facets ourselves, we believe that the upcoming long term instruments, like the \ac{cta} \cite{CTA_Introduction}, will profit from a \ac{mipod} system to keep their telescopes and telescope simulations as powerful and up to date as possible through out all the repairs and upgrades during the long expected lifetime.
%
\section*{Acknowledgments}
The important contributions from ETH Zurich grants ETH-10.08-2 and ETH-27.12-1 as well as the funding by the German BMBF (Verbundforschung Astro- und Astroteilchenphysik) and HAP (Helmoltz Alliance for Astro- particle Physics) are gratefully acknowledged. 
We are thankful for the very valuable contributions from E. Lorenz, D. Renker and G. Viertel during the early phase of the project. 
We thank the Instituto de Astrofisica de Canarias allowing us to operate the telescope at the Observatorio del Roque de los Muchachos in La Palma, the Max-Planck-Institut fuer Physik for providing us with the mount of the former HEGRA CT3 telescope, and the MAGIC collaboration for their support.
%
\section*{References}
\bibliography{references.bib}
%
\begin{acronym}
    \acro{ldm}[LDM]{Laser Distance Meter}
	\acro{fact}[FACT]{First Geiger-mode Avalanche Cherenkov Telescope}
    \acro{iact}[IACT]{Imaging Atmospheric Cherenkov Telescope}
    \acro{mipod}[MIPOD]{Mirror Position Determination}
    \acro{namod}[NAMOD]{Normalized and Asynchronous Mirror Orientation Determination}
    \acro{cta}[CTA]{Cherenkov Telescope Array}
\end{acronym}
\end{document}